\documentclass[sn-mathphys]{sn-jnl}

\usepackage{siunitx}

\theoremstyle{thmstyleone}%
\theoremstyle{thmstyletwo}%

\theoremstyle{thmstylethree}%

\unnumbered

\begin{document}

\title{Single-photon interference over 8.4 km urban atmosphere: towards testing quantum effects in curved spacetime with photons}

\author[1,2,3]{\sur{Hui-Nan Wu}}
\equalcont{These authors contributed equally to this work.}

\author[1,2,3]{\sur{Yu-Huai Li}}
\equalcont{These authors contributed equally to this work.}

\author[1,2,3]{\sur{Bo Li}}

\author[1,2,3]{\sur{Xiang You}}

\author[1,2,3]{\sur{Run-Ze Liu}}

\author[1,2,3]{\sur{Ji-Gang Ren}}

\author[1,2,3]{\sur{Juan Yin}}

\author[1,2,3]{\sur{Chao-Yang Lu}}

\author[1,2,3]{\sur{Yuan Cao}}

\author[1,2,3]{\sur{Cheng-Zhi Peng}}

\author[1,2,3]{\sur{Jian-Wei Pan}}




\affil[1]{\orgdiv{Hefei National Research Center for Physical Sciences at the Microscale and School of Physical Sciences}, \orgname{University of Science and Technology of China}, \orgaddress{\postcode{230026}, \state{Hefei}, \country{China}}}

\affil[2]{\orgdiv{Shanghai Research Center for Quantum Sciences and CAS Center for Excellence in Quantum Information and Quantum Physics}, \orgname{University of Science and Technology of China}, \orgaddress{\postcode{201315}, \state{Shanghai}, \country{China}}}

\affil[3]{\orgdiv{Hefei National Laboratory}, \orgname{University of Science and Technology of China}, \orgaddress{\postcode{230088}, \state{Hefei}, \country{China}}}

\abstract{
The emergence of quantum mechanics and general relativity has transformed our understanding of the natural world significantly. 
However, integrating these two theories presents immense challenges, and their interplay remains untested. 
Recent theoretical studies suggest that the single-photon interference covering huge space can effectively probe the interface between quantum mechanics and general relativity. 
We developed an alternative design using unbalanced Michelson interferometers to address this and validated its feasibility over an 8.4 km free-space channel. 
Using a high-brightness single-photon source based on quantum dots, we demonstrated single-photon interference along this long-distance baseline. 
We achieved a phase measurement precision of 16.2 mrad, which satisfied the measurement requirements for a gravitational redshift at the geosynchronous orbit by five times the standard deviation. 
Our results confirm the feasibility of the single-photon version of the Colella-Overhauser-Werner (COW) experiment for testing the quantum effects in curved spacetime.
}



\maketitle

The reconciliation of quantum mechanics and general relativity represents a significant challenge in modern physics \cite{carlip2001quantum, will2014confrontation}. 
However, the intersection between these two theories is typically detected under extremely difficult experimental circumstances, such as extremely tiny spatial scales ($10^{-35}$ m) \cite{pikovski2012probing}, extremely high energy scales ($10^{19}$ GeV) \cite{burgess2004quantum}, and extremely strong gravitational fields (near black holes) \cite{callan1992evanescent}.
Due to the extreme physical environment required for measurements, indirect tests have been performed, such as those involving Hawking radiation and Penrose superradiation from black holes \cite{drori2019observation, braidotti2022measurement}. 
Nevertheless, thanks to advances in technology and the efforts of theoretical physicists, a series of direct tests have recently been proposed at low-energy scales (within Earth's gravitational field) using various carriers, including optomechanics \cite{pikovski2012probing, abbott2016observation, armano2016sub}, cold atoms \cite{muller2010precision, pikovski2015universal, zych2018quantum}, atomic and optical clocks \cite{hohensee2011equivalence, delva2018gravitational, takamoto2020test, bothwell2022resolving}, and photons \cite{zych2011quantum, zych2012general, rideout2012fundamental, brodutch2015post, vallone2016interference, xu2019satellite, terno2020large, mohageg2022deep}.
Among these various carriers, photons, the fastest information carriers, are weakly coupled to the environment and suitable for large-scale quantum information experiments \cite{pan2012multiphoton, lu2022micius}. 
Tests involving massless particles, such as photons, cannot be interpreted using a Newtonian gravity framework and require a general relativity description \cite{zych2012general, mohageg2022deep}. 
Until now, all experiments that measured the influence of gravity on quantum systems are consistent with non-relativistic Newtonian gravity \cite{colella1975observation, muller2010precision}, and all tests of general relativity can be described within the framework of classical (non-quantum) physics \cite{pound1960apparent, shapiro1964fourth, hafele1972around}.
The direct experimental test for the evolution of the quantum state under curved spacetime is still lacking.
Consequently, the quantum interference of photons serves as a probe to test the interface of quantum mechanics and general relativity \cite{zych2011quantum, zych2012general, rideout2012fundamental, brodutch2015post, vallone2016interference, terno2020large, mohageg2022deep}.

\begin{figure}[h]%
\centering
\includegraphics[width=0.8\textwidth]{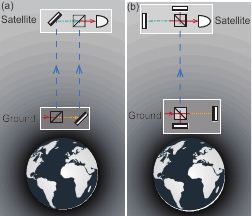}
\caption{
(a), In the MZI, single photons coherently travel along the left blue dashed and green dash-dotted paths or the right blue dashed and yellow dotted paths.
Since the two blue dashed paths traverse the same gravitational field and cancel each other out, the gravitational redshift effect will introduce the wavelength difference of single photons between the green dash-dotted and yellow dotted paths, resulting in a phase shift in single photon interference \cite{zych2011quantum, zych2012general}.
(b), To simplify the setup, two blue dashed paths are combined into one, while the yellow dotted and green dash-dotted paths become the unbalanced arm lengths of two UMI in a Franson interferometer \cite{rideout2012fundamental}.
}
\label{fig:1}
\end{figure}

As a promising avenue for future research, a satellite platform with maneuvering capabilities can potentially explore the gravitational redshift of single photons using single-photon interference \cite{rideout2012fundamental, lu2022micius}.
For example, Fig. \ref{fig:1}(a) illustrates the measurement of single-photon gravitational redshift using a Mach-Zehnder interferometer (MZI). 
Clearly, on the two paths of the MZI, the blue dashed parts pass through the same gravitational field \cite{zych2011quantum, zych2012general}. 
Meanwhile, the yellow dotted and green dash-dotted portions traverse different gravitational fields, introducing gravitational-induced phase shifts.
However, the phase noise introduced by the ground-to-satellite channel in the two arms of the MZI swamps the gravitational redshift phase, rendering the experiment impossible.
To overcome this challenge, a Franson interferometer can be utilized in Fig.\ref{fig:1}(b), consisting of two independent unbalanced Michelson interferometers (UMIs) and a shared free-space channel, equivalent to the MZI \cite{rideout2012fundamental}. 
The shared free-space channel enables the cancellation of atmospheric noise as common-mode noise. 
Assuming a satellite orbiting at an altitude of $h$ and an unbalanced arm of the interferometer of $\delta l$, the redshift phase can be expressed as
\begin{equation} 
\varphi=\frac{2\pi \delta lg}{\lambda c^2}\frac{R_eh}{R_e+h}, 
\end{equation}
where $R_e$ is the radius of the Earth, and $\lambda$ is the wavelength of the single photons \cite{rideout2012fundamental}. 
The expected observable phase shift in the experiment is 208 mrad, corresponding to a single-photon wavelength of 893.2 nm, with the satellite orbiting at an altitude of 36000 km and the UMI with an unbalanced arm length of 50 m.

Recently, Vallone et al. observed interference patterns at the single-photon level, achieving interference visibility of up to 0.67 along the ground-to-satellite channel \cite{vallone2016interference}.
In our current work, we have further concretized the scheme of the satellite-based single-photon version of the COW experiment and have explored its feasibility.
To achieve a measurement with at least 5 times the standard deviation (STD) precision, the measurement precision needs to be better than 41.6 mrad.
To confirm the feasibility of the experiment and the corresponding technology, it is necessary to perform verification experiments along the horizontal atmospheric channel.
Here, we present a demonstration of quantum interference of single photons over an 8.4 km free-space channel, achieving a multi-mode interference visibility of $\mathcal{V}=0.863\pm0.004$ and a phase measurement accuracy of $\delta\varphi=16.2$ mrad, which meets the accuracy requirement mentioned above.
To create a channel representative of the future satellite-ground channel, we chose the horizontal urban atmosphere with an equivalent thickness equal to the vertical atmosphere \cite{peng2005experimental}. 
We also developed a high-brightness quantum dot single-photon source (QDSPS) with a brightness exceeding 0.4 GHz and $g^{(2)}(0)=0.071\pm0.005$.
Our demonstration of high-precision phase measurement and high-brightness QDSPS provides a good foundation for future space-based experiments.

\begin{figure*}[h]%
\centering
\includegraphics[width=1\textwidth]{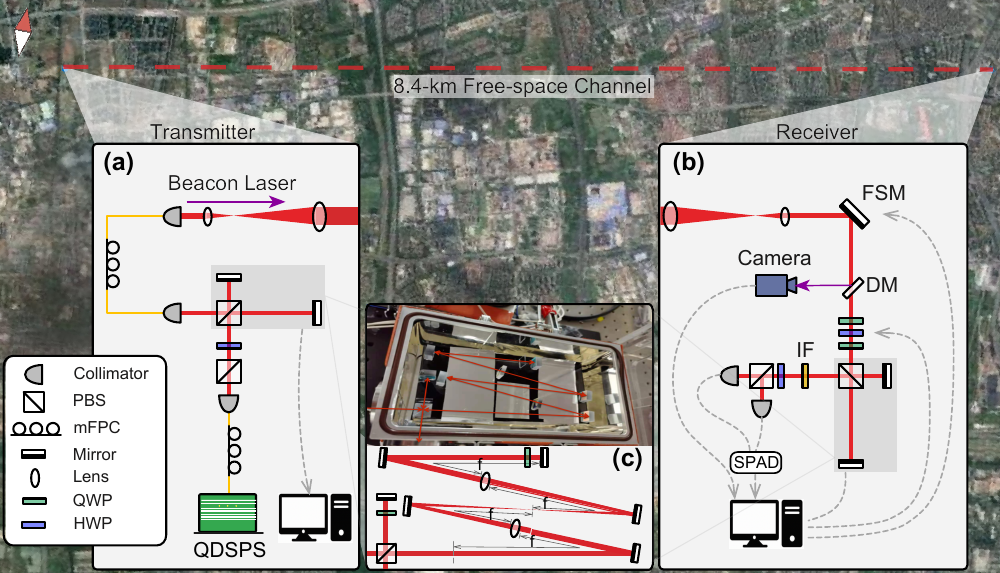}
\caption{\textbf{
Illustration of experimental setup}. 
(a), The QDSPS and UMI generate single photons in a time-bin qubit, which is crucial for achieving high-precision measurements.
\textbf{b}, The optical receiving system is mounted on a high-load pitch and yaw platform for coarse pointing. 
The telescope has an aperture of 127 mm and a beam reduction ratio of 37.5. 
The received 671-nm-wavelength laser is separated by a dichroic mirror (DM) and imaged by a camera for fine tracking of the fast steering mirror. 
A motorized rotation stage controls the three waveplates to achieve dynamic modulation of the interference phase. 
We also use a 10-nm-wide interference filter (IF) to reduce the dark count.
(c), The scheme and photo of the UMI. 
All optical elements adhere to an optical substrate that measures 160 mm in length, 80 mm in width, and 30 mm in height. 
To keep the UMI compact, we adopt a round-trip optical design and append imaging systems in the long arm of the UMI.
The red arrow represents the direction of the beam propagation in a photo of the UMI, and the interferometer is put in a vacuum chamber with a temperature controlling system.
QWP, quarter waveplate; mFPC, manual fiber polarization controller; FSM, fast steering mirror.
}
\label{fig:setup}
\end{figure*}

In our scheme, single photons are generated from a QDSPS and adjusted to state $\frac{1}{\sqrt{2}}(\vert H\rangle+\vert V\rangle)$ using a polarization beam splitter (PBS) and a half waveplate (HWP), as shown in Fig. \ref{fig:setup}(a). 
The single photons are then fed into the UMI, which transforms the state to $\vert \Psi_{t}\rangle=\frac{1}{\sqrt{2}}(\vert H\rangle\vert S\rangle+e^{i\varphi_t}\vert V\rangle\vert L\rangle)$. 
Here, $\vert L\rangle$ and $\vert S\rangle$ denote the states passing through the long and short arms of the UMI, respectively, and $\varphi_t$ is a constant internal phase of the UMI at the transmitter.
We placed the UMI in a vacuum chamber to achieve high-precision phase measurement and recorded the air pressure values using a computer, which allowed us to isolate the effects of air pressure fluctuations on our measurements.
The delay time between $\vert L\rangle$ and $\vert S\rangle$ is $\delta t\approx 4$ ns, corresponding to the UMI's arm difference of $\delta l=c\delta t\approx 1.2$ m ($c$ is the speed of light in vacuum). 
The single photons were then coupled into single-mode fiber and guided to a launch telescope.

After traveling through the free-space channel in the Shanghai metropolitan area, the single photons were captured by the receiving system, as shown in Fig. \ref{fig:setup}(b). We used adjustable waveplates to introduce a variable phase $\varphi$ between $\vert H\rangle$ and $\vert V\rangle$ state for interference visibility measuring. 
After passing through the second UMI, the state changes to $\vert \Psi_{r}\rangle=\frac{1}{\sqrt{2}}\vert L\rangle\vert S\rangle(\vert H\rangle+e^{i(\varphi_t+\varphi+\varphi_r)}\vert V\rangle)$, where $\varphi_r$ is the phase induced by the UMI at the receiver. 
Finally, we measured the photon under the Pauli operator $\sigma_x$ using an HWP and a PBS, collected photons with multi-mode fibers, with a mode field diameter of 105 \textmu m, supporting up to approximately 40000 different guided modes, and detected using single-photon avalanche detectors (SPAD).

A high-brightness, high-purity single-photon source is crucial for mitigating the significant channel attenuation that occurs in long-distance free-space communication channels. Compared to heralded single-photon sources, which generate photons probabilistically through spontaneous parametric down-conversion, solid-state single quantum emitters can produce high-quality single photons with greater efficiency and reliability.
Specifically, we employ a self-assembled In(Ga)As quantum dot embedded in a 2.5 \textmu m micropillar as a near-perfect single-photon source \cite{he2013demand, ding2016demand, deng2019quantum}. 
In our experiment, we excite the quantum dot with a narrow-linewidth continuous-wave laser, which drives the quantum dot resonantly. 
We collect the resonance fluorescence through a background-free confocal setup with an extinction ratio of about $10^7:1$. 
The count rate of single photons at 893.2 nm wavelength is more than 0.4 GHz, which reduces to about 10 kHz due to complete system attenuation. 
The purity of single photons is characterized by the second-order correlation, where the antibunching signature shows a raw $g^{(2)}(0)=0.071\pm0.005$, indicating the inherent property of resonance fluorescence.
Further details are provided in the Methods.

The ultra-stable field-widened UMI serves as the fundamental basis for this experiment. To ensure the compactness and stability of the UMI's optical path difference, we utilized ultraviolet-curing optical adhesives to integrate all optical elements onto an optical bench. These adhesives were chosen due to their low shrinkage and stress, which are important factors in preserving the stability of the UMI. Research has shown that fused silica, which has a low thermal expansion coefficient, is the optimal material for all optical elements used in this experiment \cite{li2022quantum}.

Due to the unbalanced arm length in UMI, its interference performance is significantly affected by atmospheric turbulence, including incident angle jitter and wavefront distortion. 
This effect can cause reduced interference visibility and degraded phase stability.
The spatial filter, such as single-mode fiber coupling, is generally effective, while it entails much more significant attenuation. 
An alternative solution is to incorporate an imaging system into the UMI. This approach can address the above issues while maintaining good multi-mode interference performance \cite{jin2018demonstration, jin2019genuine}.
The imaging system comprises two lenses and reduces mode-dependent path length, as illustrated in Fig. \ref{fig:setup}(c). 
Thus, it ensures high interference visibility and phase stability despite temporal and spatial photon distortion between different arms.
Since photons in the long arm pass through the imaging system twice, any spatial mode transformation becomes a nearly identical matrix that similarly affects photons in both arms. 
To evaluate the efficacy of this imaging system, we introduced an angular change of 1.6 mrad to the incident light of the UMI. 
Consequently, there was a variation of 0.04 mrad and 0.06 mm in the overlap of the beam's direction and position, respectively, for the beam passing through the long and short arms, which are much smaller than the spot diameter of 2 mm and the divergence angle of 0.55 mrad. 

This experiment aims to achieve highly accurate phase measurements by carefully isolating the interference phase from various sources of noise, as summarized in Table \ref{tab:noise}. 
Our investigation identifies six principal sources of noise, namely the photon's center wavelength, air pressure, temperature, atmospheric turbulence, shot noise, and inconsistency of SPADs.
Table \ref{tab:noise} reveals that temperature is the predominant noise source affecting long-term stability, while shot noise and inconsistency of SPADs impact short-term stability. 
Shot noise arises from the Poisson statistics of photons and can only be mitigated by increasing the number of received photons. 
The inconsistency of SPADs primarily introduces phase noise under different channel attenuations.
Furthermore, temperature and air pressure emerge as key environmental noise sources, as they introduce thermal shrinkage of interferometers and instabilities in the optical path due to variations in air pressure \cite{ciddor1996refractive}. 
To address these challenges, we placed two unbalanced Michelson interferometers (UMIs) in separate environments and implemented passive heat insulation and chamber temperature control.
Despite efforts to regulate temperature, imprecise control led to long-term drift in the interference phase of the UMI. 
Additionally, to mitigate phase noise caused by air pressure fluctuations, we maintained a vacuum environment with pressures below 0.5 Pa. 
We also categorized the effect of atmospheric turbulence on phase noise into transverse and axial components, with an integrated imaging system in UMI effectively suppressing the transverse component. Simulation results indicated that axial turbulence does not significantly contribute to the overall noise.
Furthermore, our findings highlight that the drift in the photon's center wavelength is not the primary source of noise. 
These insights shed light on the intricate interplay between noise factors and enable improved strategies for achieving high precision in interferometric phase measurements.
Further details on the experimental analysis are provided in the Methods.

\begin{table}   
\begin{center}   
\caption{Total noise analysis.} 
\begin{tabular}{cc}
     \toprule
     Noise source & Corresponding phase noise    \\
     \midrule
     Photon's center wavelength & 0.002 mrad/day \\
     Air pressure (transmitter) & 0.08 mrad \\
     Air pressure (receiver) & 0.1 mrad \\
     Temperature & 0.137 mrad/s/K \\
     Atmospheric turbulence (transverse) & 0.3 mrad \\
     Atmospheric turbulence (axial) & 0.001 mrad \\
     Shot noise & 4.3 mrad \\
     Inconsistency of SPADs & 15.6 mrad \\
     \botrule
\end{tabular}
\label{tab:noise}
\end{center}
\end{table}

\begin{figure*}[h]%
\centering
\includegraphics[width=0.9\textwidth]{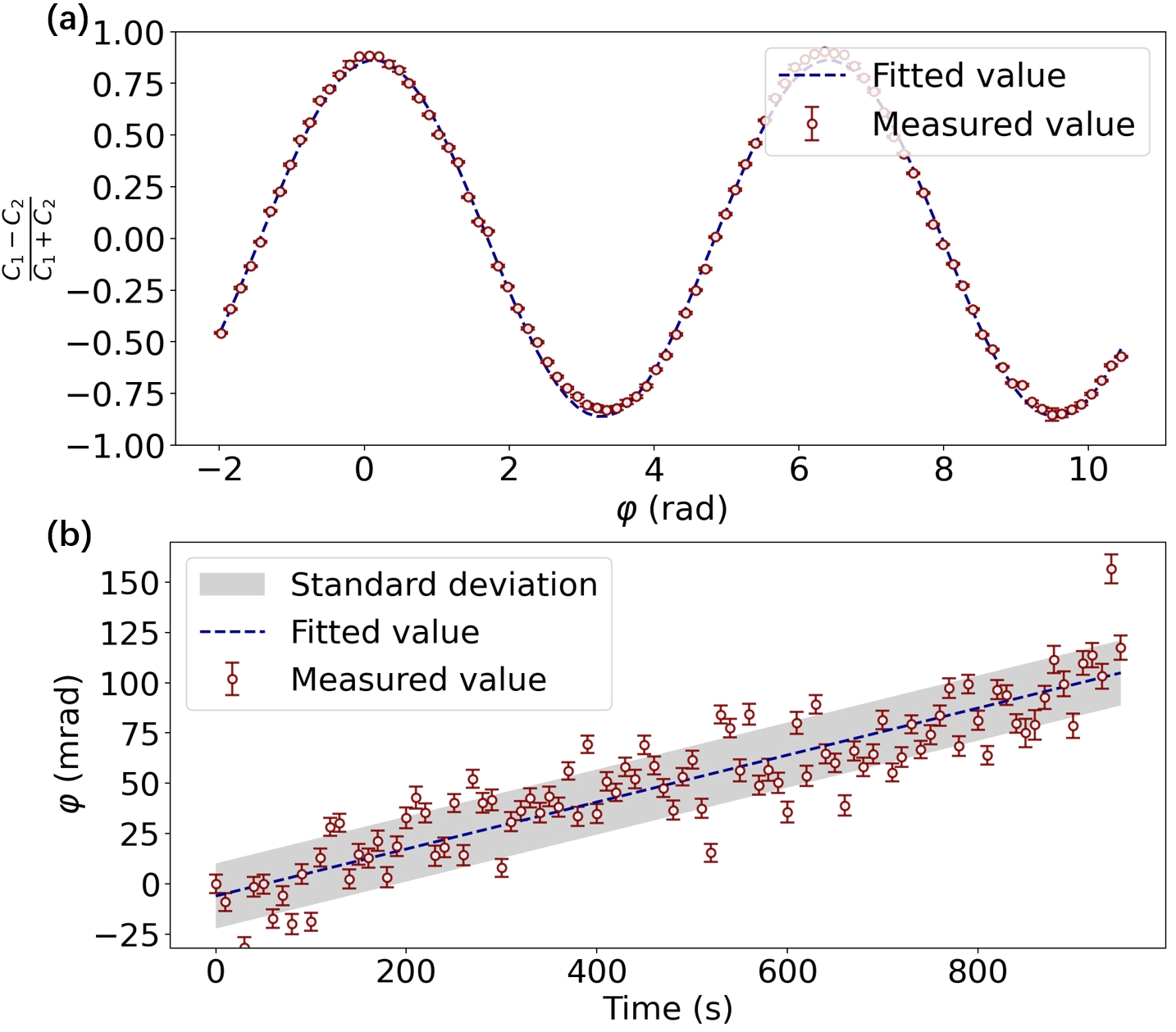}
\caption{\textbf{
Experimental results}. 
The sampling rate is set as 0.1 Hz for accumulating enough counts to weaken shot noise.
\textbf{a}, Experimental measurement of multi-mode interference visibility along free-space channel by scanning phase $\varphi$ with 2 cycles.
The scatter spots indicate the measured value, whose bars denote shot noise. The dashed line indicates the sine fitting curve, which shows the interference visibility $\mathcal{V}=0.863\pm0.004$.
\textbf{b}, Experimental measurement of phase measurement precision along free-space channel with fixed phase $\varphi$.
The red scatter spots indicate the measured value, whose bars denote shot noise.
The blue dashed line indicates the linear fitting result, which shows the long-term stability of phase measurement with slope $0.117\pm0.006$ mrad/s.
The shaded area shows the the STD of phase measurement without slope as short-term stability.
}
\label{fig:3}
\end{figure*}

We measured the multi-mode interference visibility of single-photon interference in an 8.4-km horizontal atmosphere channel located in Shanghai. 
Here, we use $\frac{C_1 - C_2}{C_1 + C_2}$ to measure interference visibility, where $C_1$ and $C_2$ represent the single-photon counts of two SPADs. 
By adjusting $\varphi$, $C_1$ can be optimized to reach its maximum value, at which point $\frac{C_1 - C_2}{C_1 + C_2}$ becomes the interference visibility $\mathcal{V}$.
In Fig. \ref{fig:3}(a), three wave plates in the receiving system introduce dynamic phase $\varphi$ for multi-mode interference. 
The measured multi-mode interference visibility $\mathcal{V}=0.863\pm0.004$, indicates the effectiveness of the imaging system against atmospheric turbulence.
By directly measuring the phase at a frequency of 0.1 Hz under fixed phase $\varphi$, we achieved a phase measurement precision of 35.8 mrad in Fig. \ref{fig:3}(b). 
However, a long-term linear phase drift was observed and found to be $0.117\pm0.006$ mrad/s, which is shown as a blue dashed line. 
Our previous noise analysis revealed that the temperature drift of the UMI causes long-term phase drift, calculated as 0.137 mrad/s/K. 
To maintain temperature stability, the air conditioner was set to a constant nighttime temperature during the experiments.
Without the long-term phase drift, the short-term stability of phase measurement precision 16.2 mrad is shown as a gray area.
Through noise analysis, we identified the inconsistency of SPADs and shot noise as the primary sources of noise. 
The shot-noise-induced phase noise was calculated to be 4.3 mrad.
The inconsistency of SPADs would inevitably introduce phase noise under various free-space channel attenuation, estimated as 15.6 mrad.
More details are provided in the Methods.

For the satellite-based experiment, it is crucial to mitigate all noise sources effectively. 
One type of noise, the long-term phase drift caused by temperature stability, can be mitigated through several approaches, such as more sophisticated active temperature control and passive thermal insulation using multiple layers \cite{armano2019temperature}. 
Additionally, employing ultra-low-expansion (ULE) glass as the optical bench for the UMI, which has a coefficient of thermal expansion two orders of magnitude lower than that of the fused silica at the specific temperature, can further mitigate the impact of temperature drift on phase stability. 
Secondly, the inconsistency of SPADs and shot noise affects short-term stability. 
To mitigate this issue, we will adopt a time-division phase measurement approach with a single SPAD.
Notably, for the satellite-based experiment, several issues warrant consideration, such as Doppler noise caused by satellite motion and so on. 
Furthermore, calibration for $\varphi_r$ and $\varphi_t$ are imperative for both UMIs located on the satellite and at the ground station.
More details are provided in the Methods.

Generally, we developed high-brightness high-purity QDSPS and high-precision single-photon phase measurement technology for future satellite-based experiments.
Therefore, we believe that our current work combined with the pioneering work of Vallone et al. \cite{vallone2016interference} have collectively demonstrated the feasibility of employing single-photon interference to measure gravitational redshift, thereby testing the interface of general relativity and quantum mechanics.
Moreover, our results identify the primary noise sources to facilitate the future use of corresponding methods to reduce noise and improve measurement precision.

\backmatter

\bmhead{Acknowledgments}
We acknowledge insightful discussions with H. Wang, J. Qin, T. Zeng and L. Huang. This work has been supported by the National Key R\&D Program of China (Grants No. 2020YFA0309803, 2020YFC2200103), the National Natural Science Foundation of China (Grants No. 11904358, 12174374, and 12274398), the Innovation Program for Quantum Science and Technology (Grants No. 2021ZD0300100), the Chinese Academy of Sciences (CAS), Shanghai Municipal Science and Technology Major Project (Grant No.2019SHZDZX01), the Natural Science Foundation of Shanghai (Grants No. 20ZR1473700), the Shanghai Rising-Star Program (Grants No. 21QA1409600) and Anhui Initiative in Quantum Information Technologies, the CAS Project for Young Scientists in Basic Research (Grant No. YSBR-046). 
Y. Cao and Y.-H. Li were supported by the Youth Innovation Promotion Association of CAS (under Grant No. 2018492, 2023475).



\newpage
\section{Methods}

\begin{figure}[h]%
\centering
\includegraphics[width=0.7\textwidth]{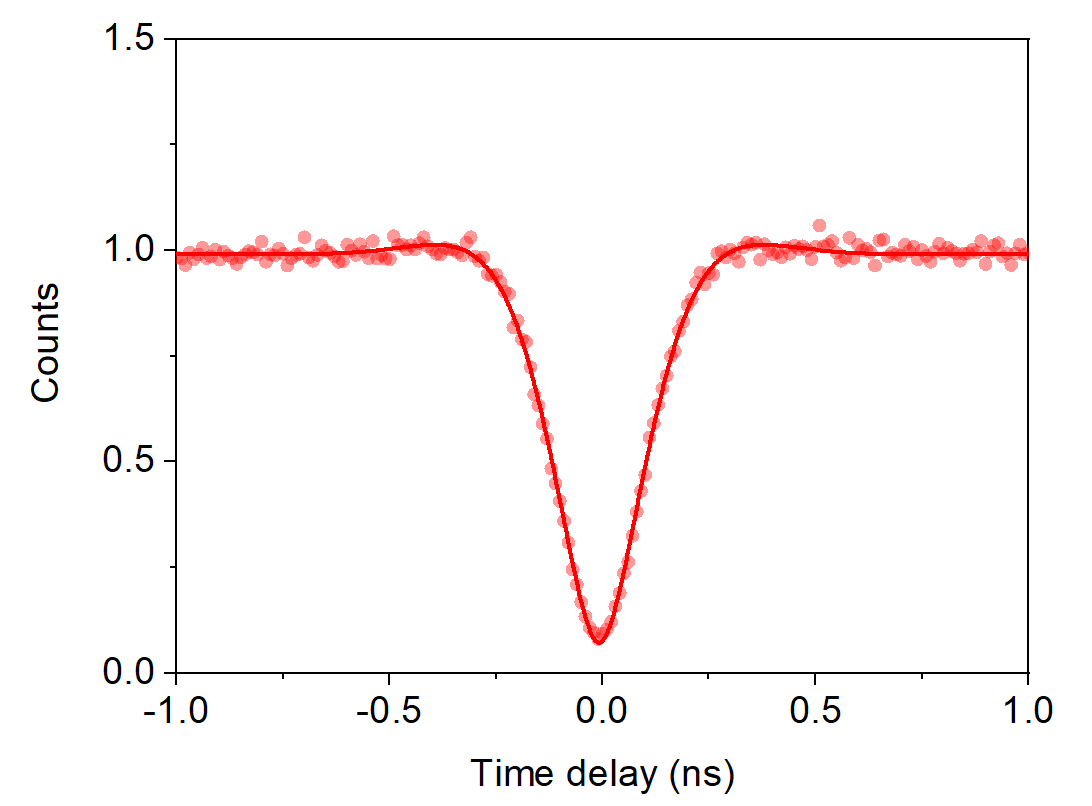}
\caption{\textbf{Second-order correlation measurement.}
The red points and lines represent the experimental and fitting data.
By sweeping the delay time, we can measure the second-order correlation with the aid of superconducting nanowire single-photon detectors.
The counts in the measurement are normalized.
}
\label{fig:g2}
\end{figure}

\textbf{Quantum Dot Single-Photon Source.}
This experiment uses a continuous-wave laser as the pump source and measures $g^{(2)}(0)$ with a superconducting nanowire single-photon detector due to its low time jitter.
With a fiber beam splitter (BS), we can measure the second-order correlation of the single-photon source to evaluate its performance.
Results show that $g^{(2)}(0)=0.071\pm0.005$ in Fig.~\ref{fig:g2}.

\textbf{Experimental noise analysis.}
Vallone et al. observed interference patterns of two temporal modes reflected by a swiftly moving satellite at the single-photon level, achieving interference visibility of up to 0.67 over channel lengths of up to 5000 km \cite{vallone2016interference}. 
Their pioneering work, utilizing a single interferometer with unbalanced arm length of 1 m and a satellite-to-ground channel, validated the feasibility of the ground-to-satellite Franson-type interferometer. 
In our current work, we have further concretized the scheme of the satellite-based single-photon version of the COW experiment and have explored its feasibility. 
By employing a true single-photon source based on the quantum dot and utilizing two independent interferometers with unbalanced arm length of 1.2 m to simulate the satellite-based experiment, we achieved an interference visibility of 0.863 and a phase measurement accuracy of 16.2 mrad along an 8.4 km horizontal atmospheric channel. 
Additionally, we conducted a comprehensive noise analysis on phase measurements, further showcasing the potential of the satellite-based experiment.

Due to the inconsistency of two interferometers, photons' center wavelength drift will induce phase uncertainty in measurement.
For $\lambda$-wavelength photons with $\Delta\lambda$-center-wavelength drift will introduce phase noise as 
\begin{equation}
\Delta\varphi_{\rm cen}=\frac{2\pi\delta l\Delta\lambda}{\lambda^2}.
\end{equation}
The arm difference of two unbalanced Michelson interferometers (UMIs) is estimated as 8.25 \textmu m, and the variance of quantum dot single-photon source's (QDSPS) center frequency is 10 MHz/day approximately, so there will be 0.002 mrad phase noise in the measurement.
This is not the primary source of noise in the experimental setup.

Meanwhile, the air pressure variance will induce variance of the index of refraction, and therefore introduce phase noise.
So we put UMIs in vacuum chambers, and photons will pass in and out of the chamber through sealed optical windows.
For unbalanced arm 1.2 m in our setup, we have measured the equivalent phase standard deviation (STD) induced by air pressure, which is 0.1 mrad and 0.08 mrad in the transmitter and receiver with the aid of vacuum meters, respectively~\cite{ciddor1996refractive}.

The thermal expansion of fused silica can also introduce phase noise, as heat transfer is known to occur through conduction, convection, and radiation. 
However, by utilizing vacuum technology, the influence of convective heat transfer can be reduced. We mounted interferometers on three hollow thin polyacrylamide-made pillars and applied high-reflectance silver film around UMI to passively slow down the speed of heat conduction and radiation, respectively.
Theoretically, we calculated temperature-induced phase drift speed as
\begin{equation}
\dot{\varphi}_{\rm temp}=[\frac{2\pi k\alpha A_1\delta l}{Cmh\lambda}+\frac{2\pi C_0A_2}{\frac{2}{\epsilon_1}-1}(\frac{T}{100})^4\frac{4\Delta T}{T}\frac{\alpha\delta l}{Cm\lambda}]\Delta T,
\end{equation}
where $k$, $A_1$, and $h$ denote the pillars' thermal conductivity, cross-sectional area, and length.
$\alpha$, $C$, and $m$ denote the UMI's coefficients of thermal expansion, specific heat capacity, and mass.
$T$ and $\Delta T$ denote environmental temperature and the difference between the environment and UMI.
$A_2$ and $\epsilon_1$ denote the surface area of UMIs and their emissivity.
$C_0$ is a constant in Kirhoff's radiation law.
With the aid of air conditioning, we were able to maintain temperature control within a range of $\pm$1 K inside our experimental setup. 
Within the confines of our configuration, it was possible to evaluate temperature-induced phase drift as $\dot{\varphi}_{\rm temp}=0.137\times\Delta T$ mrad/s/K. 
This suggests that even a difference of 1 K between the surroundings and interferometers can cause a drift in phase of approximately 0.137 mrad/s. 
We observed a similar phenomenon in our experiment, with a measured value of $0.117\pm0.006$ mrad/s closely matching our expected outcome.
Given our ability to regulate temperature, we identified temperature-induced phase drift as the primary source of long-term drift in the interference phase. 

The noise generated from atmospheric turbulence can be divided into transverse and axial.
The former causes wavefront distortion and fluctuations of the angle of incidence (AOI), while the latter results in a variation in the arrival time of photons, which is also known as atmospheric phase noise.
The phase noise resulting from AOI fluctuation can typically be corrected using UMI's imaging systems. 
After closed-loop tracking by the fast steering mirror, the STD of AOI amounts to 62 \textmu rad. 
There would be a 0.3 mrad phase noise in the experiment, which is approximately 183 times smaller than that without the imaging system in UMI.
Another approach to mitigating phase noise from transverse turbulence is through single-mode fiber coupling. 
The results indicate that the phase stability is almost equivalent to the previous one, which suggests that the imaging system is highly effective at eliminating noise.
Regarding axial turbulence, it usually causes variations in the arrival time of photons due to frozen turbulence with a constant wind speed, $v$, perpendicular to the beam direction. 
The theoretical description of atmospheric phase noise is commonly based on the Kolmogorov spectrum
\begin{equation}
S_\varphi(f)=0.016k^2LC_n^2v^{5/3}f^{-8/3},
\end{equation}
where $C_n^2$ is the turbulence structure constant, $k$ is the wave number, and L is atmospheric channel length. 
We measured Fried constant as 53 mm with a 671-nm beacon laser, so $C_n^2$ can be calculated as \SI{4.5e-16}{m^{-2/3}}.
In our experimental setup, the time delay of UMI amounts to 4 ns, which indicates that we only need to consider phase noise at 0.25 GHz. 
Assuming a wind speed of 5 m/s, the Kolmogorov spectrum can be calculated as \SI{1.7e-21}{rad^2/Hz}, resulting in phase noise as 1 \textmu rad. 
Based on the theoretical calculation and simulations conducted, it can be concluded that atmospheric phase noise is not the primary source of noise in our experiment.

The statistics of photons will give rise to shot noise, which is attributed to the random absorption events of photons by a single-photon avalanche detector (SPAD). 
In the experiment, we can only accumulate sufficient events to reduce shot noise for a single measurement.
Suppose $C_1$ and $C_2$ denote the counts obtained from two detectors, the shot noise is given by the square root of $C_1$ and $C_2$, respectively.
Therefore, the shot noise can be written as
\begin{equation}
    \Delta\varphi_{\rm sn}=2\frac{\sqrt{C_1^2C_2+C_1C_2^2}}{\mathcal{V}(C_1+C_2)^2\sqrt{1-\frac{(C_1-C_2)^2}{\mathcal{V}^2(C_1+C_2)^2}}}.
\end{equation}

\begin{table}   
\begin{center}   
\caption{Noise comparison in the single-SPAD setup.} 
\begin{tabular}{ccc}
     \toprule
     Average photon counts & Measuring noise  & Shot noise  \\
     \midrule
     \num{1.1e4} & 9.1 mrad & 9.5 mrad \\
     \num{3.6e4} & 5.2 mrad & 5.2 mrad \\
     \num{1.1e5} & 2.8 mrad & 3.0 mrad \\
     \num{3.6e5} & 1.3 mrad & 1.7 mrad \\
     \num{1.1e6} & 0.8 mrad & 1.0 mrad \\
     \botrule
\end{tabular}
\label{tab:det1}
\end{center}
\end{table}

\begin{table}   
\begin{center}   
\caption{Noise comparison in the dual-SPAD setup.} 
\begin{tabular}{ccc}
     \toprule
     Average photon counts & Measuring noise  & Shot noise  \\
     \midrule
     \num{1.1e4} & 9.6 mrad & 9.8 mrad \\
     \num{3.1e4} & 5.6 mrad & 5.6 mrad \\
     \num{1.1e5} & 3.0 mrad & 3.1 mrad \\
     \num{3.2e5} & 1.9 mrad & 1.8 mrad \\
     \num{1.1e6} & 1.2 mrad & 1.0 mrad \\
     \botrule
\end{tabular}
\label{tab:det2}
\end{center}
\end{table}

\begin{figure}[h]%
\centering
\includegraphics[width=0.7\textwidth]{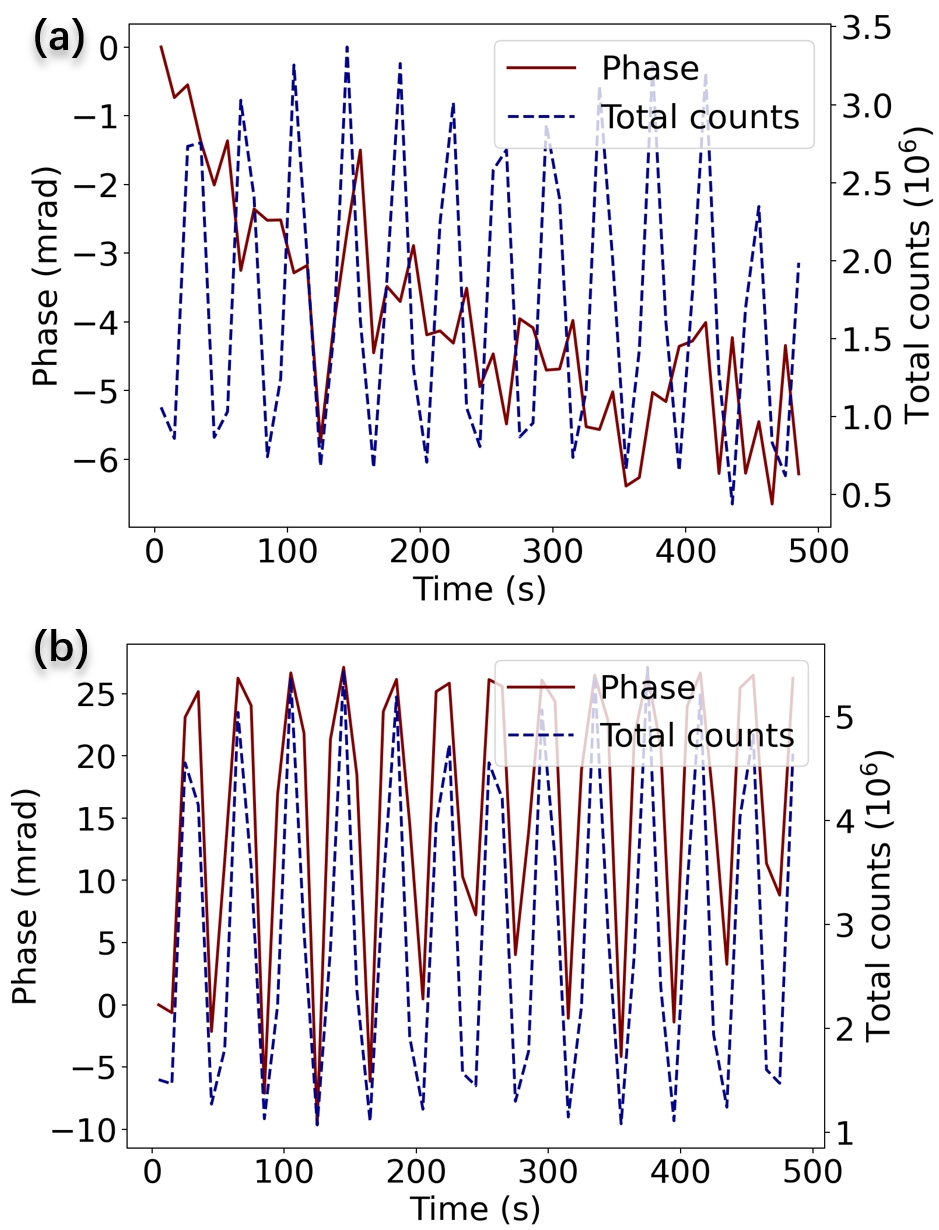}
\caption{\textbf{Noise test for inconsistency of SPADs.}
\textbf{a}, Correlation between the measuring phase and total photon counts in the single-SPAD setup.
\textbf{b}, Correlation between the measuring phase and total photon counts in the dual-SPAD setup.
The sampling rate is 0.1 Hz.
}
\label{fig:det}
\end{figure}

The inconsistency of SPADs can significantly impact the precision of our phase measurement. 
To achieve high-precision phase measurement, we must ensure a high-precision ratio of photon count.
In a dual-SPAD setup, we use an attenuated laser as the photon source and simulate varying channel attenuation in the lab using a variable neutral density filter driven by a motorized rotation stage, which is similar to the outfield experimental setup. 
In contrast, in a single-SPAD setup, we use an attenuated pulsed laser as the photon source, with a pulse width and repetition rate of 3 ps and 75.9 MHz, respectively. 
The attenuation simulation is identical to that used in the dual-SPAD setup.
After splitting the photons with fiber BSs, we detect trigger photons for synchronization and split the remaining ones into two fibers with a BS. 
Using a single-mode fiber, we introduce a 3.1 ns time delay before combining the signals and distinguishing them in the time domain with a time-to-digital converter.
At a fixed attenuation, the measurement noise is predominantly due to shot noise in both setups, regardless of the average photon count, as shown in Table~\ref{tab:det1} and Table~\ref{tab:det2}.
In the single-SPAD setup, inconsistency between different SPADs can be canceled compared to those in the dual-SPAD setup when calculating the interference phase using the photon count ratio in adjacent time windows. 
At a fixed attenuation, the measurement noise is predominantly due to shot noise in both setups, regardless of the average photon count, as shown in Table~\ref{tab:det1} and Table~\ref{tab:det2}. 
Thus, system noise is not the primary source of noise. 
It is worth noting that any phase noise introduced by the inconsistency of SPADs would be small compared to shot noise, indicating that it is not the primary source of noise.
To simulate atmospheric effects, we introduce varying attenuation with an amplitude of approximately 7 dB and a period of 38 seconds in both the single-SPAD and dual-SPAD setups. 
The results are shown in FIG.~\ref{fig:det}, in which the blue dashed line and the red solid line represent total counts and equivalent phase, respectively.
In the single-SPAD setup, as the counts change, there is no corresponding variation in the phase in Fig.~\ref{fig:det}(a).
A phase drift of 9.2 \textmu rad/s is observed, possibly due to the stability of the fiber BS.
The STD of the phase is 1.6 mrad in the single-SPAD setup, and shot noise is measured as 0.8 mrad. 
However, the results indicate a significant correlation between the variation in the measuring phase and the total photon count in the dual-SPAD setup, as shown in Fig.~\ref{fig:det}(b). 
The STD of the measuring phase is 11.4 mrad, with a shot noise of 0.9 mrad. 
We evaluate the variation of channel attenuation by the ratio of the STD to the mean of the counting rate. 
In this test, this value is measured as 0.52. 
However, in a realistic atmospheric environment, this value is 0.71, resulting in an amplification of the corresponding noise to 15.6 mrad.
Thus, inconsistencies between different SPADs could result in a phase noise of 15.6 mrad for the experiment.

In our comprehensive analysis, we systematically investigate and classify the various sources of phase noise into two distinct categories: those that stem from the unbalanced arm of the UMI, including center wavelength stability, air pressure stability, temperature stability, and atmospheric phase noise; and those unrelated to the unbalanced arm, such as AOI jitter, shot noise, and SPADs' inconsistency.
By extending the unbalanced arm to a length of 50 m, it becomes feasible to suppress the phase noise associated with the unbalanced arm to an approximate level of 1 mrad. 
This can be achieved by implementing techniques such as enhanced vacuum conditions and replacing the interferometer substrate with ultra-low-expansion (ULE) glass, among other methods.
The noise not arising from the unbalanced arm primarily originates from SPADs' inconsistency, contributing to an estimated 15.6 mrad.
Remarkably, considering a gravitational redshift of 208 mrad, our measurement demonstrates the potential for achieving a precision five times greater than the STD.

\textbf{Analysis for the satellite-based experiment.}
In the satellite-based experiment, various challenges remain, including orbital stability, tidal movements, and channel efficiency. 
This section delves into an analysis of these issues.

The UMI in the ground station coherently splits single photons with a delay of $\delta l/c$. 
Consequently, due to this time delay, two superpositions of single photons experience different velocities relative to the ground station upon entering the satellite.
If we assume a radial velocity $v_r$ for the satellite, its motion within a given time delay introduces an additional phase, given by
\begin{equation}
\varphi_{dop}=\frac{2\pi\delta l}{c\lambda} v_r,
\end{equation}
which has been demonstrated by Vallone et al. \cite{vallone2016interference}.
We can achieve a radial velocity measurement accuracy of 1 mm/s through precise orbit determination techniques, corresponding to an introduced Doppler phase noise of 1.2 mrad. 
In the satellite-based experiment, real-time phase adjustment with waveplates can eliminate this noise component.

Regarding tidal movements, preliminary estimates indicate that contributions from the lunar and solar tidal potentials, may be on the order of several $10^{-17}$ \cite{qin2019relativistic}, significantly smaller than gravitational redshift effects. 
Consequently, this factor is not considered in the current experiment.

\begin{table}   
\begin{center}   
\caption{Efficiency analysis for the satellite-based experiment.} 
\begin{tabular}{cc}
     \toprule
     Item & efficiency (dB) \\
     \midrule
     UMI (satellite) & 1\\
     Telescope (satellite) & 2 \\
     Atmospheric transmittance & 0.5 \\
     Geometric efficiency & 59 \\
     Telescope (ground station) & 2 \\
     Multi-mode coupling & 1\\
     UMI (ground station) & 1\\
     SNSPD & 1 \\
     Total & 67.5 \\
     \botrule
\end{tabular}
\label{tab:eff}
\end{center}
\end{table}

Moreover, channel efficiency for the satellite-based experiment has been evaluated and is detailed in Table \ref{tab:eff}.
Assuming a satellite telescope aperture of 1.2 m at an orbital altitude of 35786 km and a beam divergence angle of 30 \textmu rad on the ground station (considering beam wandering from atmospheric turbulence), the geometric efficiency is estimated at 59 dB. 
We plan to use superconducting nanowire single-photon detectors (SNSPDs) instead of SPADs to boost system efficiency. 
After our actual assessment, the total efficiency of the system is estimated at 67.5 dB. 
Therefore, a shot noise of 4.3 mrad requires an acquisition time of approximately 0.28 hours.
In our experiment, such a shot noise requires an acquisition time of 10 seconds. 
We collected photon counts for a total of 0.26 hours in the experiment, which is consistent with the acquisition time of the satellite-based experiment.
If we further suppress the calibrated noise source in the experiment, it is evident that our results underscore the feasibility and promise of the satellite-based experiment.

\textbf{Interferometer calibration.}
The ability to calibrate the phase stability of a single interferometer is one of the crucial aspects of the satellite-based experiment, thereby enabling precise measurements of the gravitational redshift effect on single photons.

For the UMI on the ground station, we primarily employ two methods for calibrating unbalanced arm length. 
The first involves using homodyne interference with the ultrastable laser locked to an optical atomic clock \cite{bothwell2022resolving, li2024strontium} in a UMI to calibrate the phase stability.
Assuming the wavelength of the ultrastable laser is 900 nm, typically with a linewidth below 1 kHz, this implies that its coherence length can reach 300 km, which far exceeds the 50 m unbalanced arm length of the interferometer, allowing it to form interference patterns.
Moreover, if the ultrastable laser's central frequency stability is better than 1 kHz, then the phase noise introduced by frequency noise would be less than 1 mrad, meeting the requirements for calibration precision.
While this method offers high precision, its unambiguity range is limited to half a wavelength level due to the constraints of homodyne interference.
Secondly, we employ dual-comb ranging to extend the unambiguity range.
This method involves using two optical frequency combs (OFC) with a slight difference in repetition rates. 
By directing the pulses from the probe OFC into the interferometer to obtain two trains of femtosecond pulses through the long and short arms, probe pulses then will beat with the pulses from the local OFC.
With each sequential pair of probe and local pulses, the relative pulse timing shifts slightly, the cross-correlation is formed in the interference signal. 
For sufficiently stable femtosecond lasers, this cross-correlation contains precise information about the timing between the pulses through the long arm and short arm; the pulse envelopes provide a time-of-flight range measurement, while the carrier phase under the pulse provides an interferometric range measurement.
There has been remarkable progress on dual-comb ranging that can improve the precision of displacement measurement to several nanometers \cite{coddington2009rapid, lee2010time, trocha2018ultrafast}.
For example, Coddington et al. demonstrated an absolute distance measurement with precision below 3 nm at an acquisition time of 0.5 s using such method \cite{coddington2009rapid}, and Lee et al. achieved a precision of 1.1 nm at 1 s \cite{lee2010time}, which met the requirement for the satellite-based experiment.
By combining the above two methods, we can effectively calibrate the phase stability of the interferometer.

\begin{figure}[htbp]%
\centering
\includegraphics[width=0.7\linewidth]{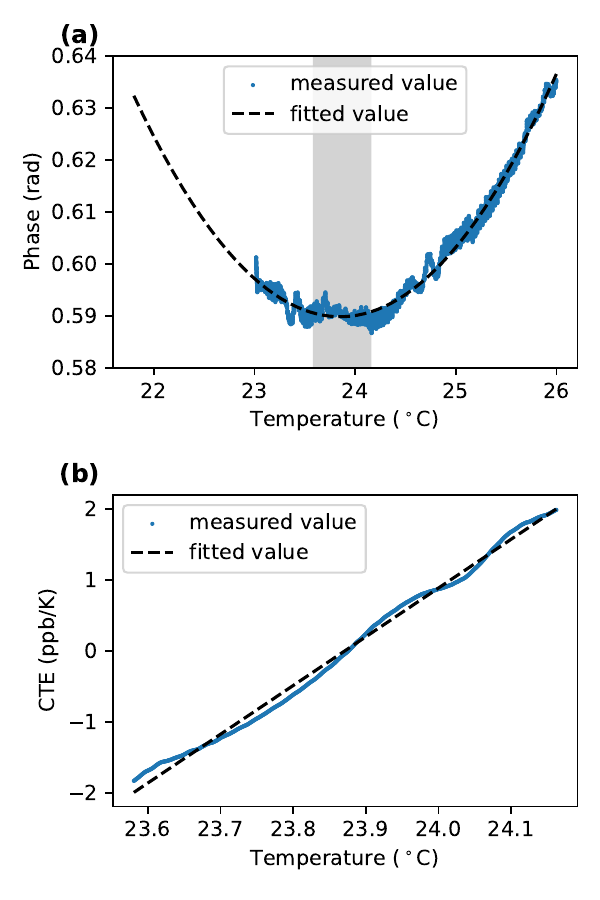}
\caption{\textbf{CTE test of the ULE glass.}
\textbf{a}, The interference phase varies with temperature in a quadratic polynomial form. 
The gray area indicates the optimal temperature range for ULE glass.
\textbf{b}, Corresponding to the optimal temperature range, the CTE of ULE glass changes linearly with temperature.
The scatter blue spots indicate the measured value, and the dashed red line indicates the fitting value.
}
\label{fig:cte}
\end{figure}

For the UMI on the satellite, when conditions permit, we can consider employing the same calibration method as above. 
Considering the limited resources available on a satellite, we can conduct the proposed experiment on an elliptical orbit, such as with a perigee of 10000 km and an apogee of 20000 km. 
In this scenario, the gravitational redshift-induced phase difference between the two orbital points would be 36.2 mrad. 
Considering the orbit above, where the satellite takes approximately 5 hours to move from the perigee to the apogee, we must ensure the phase of the UMI remains constant. 
In our experiment, we have identified that the long-term phase stability of the interferometer predominantly relies on the UMI's temperature. 
The thermal expansion coefficient (CTE) of ULE glass varies linearly with temperature and reaches zero at a specific temperature, leading to the phase changing with temperature in a quadratic polynomial form.
To verify the performance of ULE glass, we completed a Michelson interferometer with an unbalanced arm length of 0.8 m on the ULE glass and used a 1550-nm ultrastable laser as the light source to eliminate phase noise caused by fluctuation of the central wavelength.
By placing the interferometer in a vacuum chamber and heating it from 23 $^\circ$C, we plotted the curve of phase variation with temperature in Fig. \ref{fig:cte}(a).
The obtained coefficient of determination $R^2$ is 0.99, indicating the consistency between the experimental results and expectations.
The gray area represents the optimal operating range for ULE glass, and we obtained the the curve of CTE variation with temperature, as shown in Fig. \ref{fig:cte}(b).
Considering the current technology, achieving a temperature control accuracy of 0.2 $^\circ$C for the payload is feasible.
The results indicate that the CTE of ULE glass reached up to 1.4 ppb/K within 23.87$\pm$0.2 $^\circ$C.
Considering the CTE of fused silica is approximately 550 ppb/K, the ULE glass can effectively suppress the phase stability variation approximately 393 times under the same temperature control conditions, which meets the requirements for the satellite-based experiment.
Therefore, we can enforce precise thermal management for the interferometer, employ ULE glass with reduced thermal expansion coefficient as the substrate of the UMI, and continuously monitor the interferometer's temperature in real-time to calibrate the phase indirectly. 
Therefore, the phase precision of 16.2 mrad has already met the requirements for measuring gravitational redshift, which is demonstrated in our experiment. We can adopt a time-division phase measurement approach with a single SPAD to further enhance the precision of phase measurements.
Through the methods above, we can achieve calibration of the phases $\varphi_r$ and $\varphi_t$ in the satellite-based experiment.

\end{document}